\documentclass[twocolumn,superscriptaddress]{revtex4}
\usepackage{graphicx}
\usepackage{amsmath}
\usepackage{amsfonts}
\usepackage{amssymb}
\usepackage{mathrsfs}

\def\openone{\leavevmode\hbox{\small1\kern-3.8pt\normalsize1}}
\def\N{\leavevmode\hbox{ Z \kern-8 pt\normalsize{Z}}}
\def\openone{\leavevmode\hbox{\small1\kern-3.8pt\normalsize1}}
\def\openJ{\leavevmode\hbox{J \kern-9.5pt\normalsize J}}
\def\openS{\leavevmode\hbox{ S \kern-9.3pt\normalsize S}}
\newcommand{\bb}{\begin{equation}}
\newcommand{\ee}{\end{equation}}
\newcommand{\eqb}{\begin{eqnarray}}
\newcommand{\eqf}{\end{eqnarray}}

\usepackage{color}

\begin{document}

\title{Collisionless Magnetic Reconnection in Curved Spacetime \\ and the Effect of Black Hole Rotation}
\author{Luca Comisso}
\email{lcomisso@princeton.edu}
\affiliation{Department of Astrophysical Sciences and Princeton Plasma Physics Laboratory, Princeton University, Princeton, NJ 08544, USA.}
\author{Felipe A. Asenjo}
\email{felipe.asenjo@uai.cl}
\affiliation{Facultad de Ingenier\'{\i}a y Ciencias, Universidad Adolfo Ib\'a\~nez, Santiago 7941169, Chile.}


\begin{abstract}

Magnetic reconnection in curved spacetime is studied by adopting a general relativistic magnetohydrodynamic model that retains collisionless effects for both electron-ion and pair plasmas. A simple generalization of the standard Sweet-Parker model allows us to obtain the first order effects of the gravitational field of a rotating black hole. It is shown that the black hole rotation acts as to increase the length of azimuthal reconnection layers, per se leading to a decrease of the reconnection rate. However, when coupled to collisionless thermal-inertial effects, the net reconnection rate is enhanced with respect to what would happen in a purely collisional plasma due to a broadening of the reconnection layer. These findings identify an underlying interaction between gravity and collisionless magnetic reconnection in the vicinity of compact objects. 
 
\end{abstract}


\maketitle

\section{Introduction}

Magnetic reconnection is a fundamental process whereby the connectivity of the magnetic field lines   \cite{Newcomb,pegoraroEPJ,asenjoComissoCon,AsenjoComissoPRD17} is modified due to the presence of a localized diffusion region. This results in a rapid conversion of magnetic energy into kinetic and thermal energy. As such, magnetic reconnection is believed to power some of the most important and spectacular astrophysical phenomena in the Universe such as stellar flares, gamma-ray flares in blazar jets, and non-thermal emissions from active galactic nuclei \citep{Kulsrud_2005,KZ15}.

Although the theory of magnetic reconnection has been mainly focused in the non-relativistic regime \cite{YKJ_2010}, in recent years there has been a growing body of studies aimed towards understanding magnetic reconnection in magnetically dominated environments, where special relativistic effects must be considered \cite{Uzdensky2011,Hoshino2012}. Indeed, in these situations the magnetic energy density exceeds the relativistic enthalpy density, implying that the Alfv\'en speed approaches the speed of light. This motivated the problem of the special relativistic generalization of the collisional Sweet-Parker and Petschek reconnection models, which was approached for the first time by Blackman and Field \cite{Blackman_1994}. They argued that the inflow velocity of the reconnecting magnetic field would be enhanced to ultra-relativistic speeds because of Lorentz contraction. The same conclusion was drawn by Lyutikov and Uzdensky \cite{LyutUzd_2003}, who performed a similar analysis focused on the Sweet-Parker reconnection model. 

However, few years later Lyubarsky \cite{lyu} showed that in the collisional Sweet-Parker regime, the thermal pressure within the reconnection layer constrains the outflow velocity to be mildly-relativistic, therefore limiting the inflow velocity to remain sub-relativistic.  In the collisional Petschek scenario, he found that the outflow velocity becomes ultra-relativistic, but the reconnection velocity remains sub-relativistic because the Lorentz contraction is compensated by a decrease of the angle between the Petschek shocks as the magnetization parameter increases. In consideration of the fact that resistive relativistic magnetohydrodynamic simulations confirmed these predictions \cite{Watanabe2006,Zenitani_2010,Takahashi_2011}, Lyubarsky's theory became the standard theory for collisional reconnection in the special relativistic regime.

On the other hand, simulations of relativistic reconnection with two-fluid or kinetic models showed reconnection rates higher than those predicted by Lyubarsky's analysis \cite{Zenitani2008,Zenitani2009,Bessho2012,Sironi2014,Guo2014,Liu2015}. Therefore, a further generalization to account for collisionless effects in the special relativistic regime was needed. This was done in a fairly recent Letter \cite{luca1}, where it was shown that thermal-inertial effects played an essential role in increasing the reconnection rate with respect to the purely resistive case discussed by Lyubarsky. Indeed, thermal-inertial effects were found to modify the Sweet-Parker and Petschek relativistic scenarios by causing a broadening of the reconnection layer that is capable of supporting a larger inflow velocity of the reconnecting magnetic field.

While special relativistic effects on the magnetic reconnection process are becoming increasingly recognized, general relativistic effects are far less investigated. Several studies have predicted the formation of reconnection layers in the vicinity of black holes \cite{Koide2006,Karas2009,Penna2010,LyutMcK_2011,McKinney2012,Karas2012,Ball2017}, where the effects of the spacetime curvature can be important. However, a detailed investigation of these reconnection layers by means of numerical simulations has not been possible so far, due to the stringent requirements on the spatial and temporal resolutions of typical reconnection processes in this setting. On the other hand, an advance in our theoretical understanding of magnetic reconnection in curved spacetime has been recently obtained by generalizing the collisional Sweet-Parker reconnection model in Kerr spacetime \cite{asenjoGRLuca}.
This approach allowed us to obtain a first estimation of the effects of the gravitational field on the magnetic reconnection process in the collisional regime.

In this paper, we take a step forward by considering also the contribution of collisionless effects. Indeed, collisionless effects are expected to be more important than collisional ones in plasmas surrounding black holes (e.g., Ref. \cite{Quat2002}). To perform this analysis, we adopt a set of equations for a general relativistic magnetohydrodynamic (GRMHD) model that retains two-fluid effects. Both electron-ion and pair plasmas can be described within this model, and we obtain the reconnection rate and other properties of the reconnection layer for both cases. We find that in the collisionless regime there is an interaction between the thermal-inertial and the spacetime curvature effects. The combination of these effects lead to a broadening of the reconnection layer and a net reconnection rate that is larger than what would have been predicted considering a purely collisional case.

The outline of the paper is as follows. The model equations are presented in Sec. II, while the spacetime of the rotating black hole and the configuration of the reconnection layer are given in Sec. III. We derive the properties of the reconnection layer and the reconnection rate in Sec. IV. Finally, the most relevant results are summarized in Sec. V.

\section{Model Equations}

We commence our study from the covariant form of the generalized GRMHD equations derived by Koide in Ref.~\cite{Koide_2010}. 
They include the usual continuity equation
\begin{equation}\label{contAp}
\nabla_\nu\left(n U^\nu\right)=0\, ,
\end{equation}
where  $n$ is  the proper particle number density of the magnetohydrodynamic plasma,  $U^\mu$ is its four-velocity, and  $\nabla_\nu$ denotes the covariant derivative associated with the spacetime metric $g_{\mu\nu}$. 
The generalized version of the momentum equation is 
\begin{equation}\label{MomAp}
\nabla_\nu\left[h \left(U^\mu U^\nu+\frac{\xi}{4n^2e^2}J^\mu J^\nu\right)\right]=-\nabla^\mu p+J_\nu F^{\mu\nu}\, ,
\end{equation}
where $e$ is the electron charge, $h = {n^2}({h_ + }/n_ + ^2 + {h_ - }/n_ - ^2)$ is the proper enthalpy density (subscripts $+$ and $-$ are used to indicate the positively and negatively charged fluids), $p=p_+ + p_-$ is the proper pressure, $J^\mu$ is the four-current density,  $F^{\mu\nu}$ is the electromagnetic field tensor, and finally
\begin{equation}
\xi=1-(\Delta\mu)^2 \, , 
\end{equation}
with
\begin{equation}
\Delta\mu=\frac{m_+-m_-}{m_+ + m_-} \, 
\end{equation}
denoting the normalized mass difference of the positively and negatively charged particles. 
Notice that $\xi$ is a constant in agreement with the covariant transformation of the momentum equation ($\xi\approx 4m_-/m_+$ for an electron-ion plasma, while $\xi=1$ for a pair plasma). 
Then, the plasma dynamics is completed by the generalized Ohm's law
\begin{eqnarray}\label{OhmAp}
&&\frac{1}{4en}\nabla_\nu\left[\frac{\xi h}{ne}\left(U^\mu J^\nu+J^\mu U^\nu-\frac{\Delta\mu}{ne}J^\mu J^\nu\right)\right]\nonumber\\
&&=\frac{1}{2ne}\nabla^\mu (\Delta\mu\, p- \Delta p)+ \left(U_\nu-\frac{\Delta\mu}{ne}J_\nu\right) F^{\mu\nu}\nonumber\\
&&-\eta \left[ {J^\mu - \rho '_e  (1 + \Theta) U^\mu} \right] \, ,
\end{eqnarray}
where $\Delta p = p_+ - p_-$ is the pressure difference between the fluids, $\rho '_e = -U_\nu J^\nu$ is the charge density observed by the local center-of-mass frame, $\Theta$ is the thermal energy excange rate between the two fluids \cite{Koide_2010}, and $\eta$ is the electrical resistivity, which is considered as a phenomenological parameter in this model.
Finally, the system is completed by Maxwell's equations
\begin{equation}\label{MaxGeneralCurvedAp}
\nabla_\nu F^{\mu\nu}=J^\mu \, , \qquad \nabla_\nu F^{*\mu\nu}=0 \, ,
\end{equation}
where $F^{*\mu\nu}$ is the dual of the electromagnetic field tensor.

The derivation of this system of generalized GRMHD equations that retain two-fluid effects assumes ${n_+} \approx {n_-}$ and $\Delta h = m{n^2}({h_ + }/{m_ + }n_ + ^2 - {h_-}/{m_-}n_-^2)/2 \ll h$. We recall that the terms proportional to $h/(ne)^2$ in the left-hand side of Eqs.~\eqref{MomAp} and \eqref{OhmAp} are thermal-inertial terms, while the first two terms and the fourth term in the right-hand side of Eq. \eqref{OhmAp} are Hall terms. For definiteness, in this work we focus on the investigation of magnetic reconnection in the thermal-inertial regime, which correspond to the situation in which the thermal-inertial terms are larger than the Hall terms \cite{kimura14,Lingam15,Lingam16,Kawa17}. For an electron-ion plasma, assuming that the Hall terms are of the same order, the thermal-inertial regime can be achieved if the condition  
\begin{equation}\label{}
\dfrac{{\Delta \mu J B}}{\left({\dfrac{{\xi h}}{{n e \ell}}}\right)U J} \sim {\Omega _e}\tau  \ll 1 \, 
\end{equation}
is satisfied. Here, $\Omega _e$ is the electron gyro-frequency, while $\ell$ and $\tau$ are the characteristic length and time scales of current change. On the other hand, for a pair plasma, the thermal-inertial regime is naturally satisfied because $\Delta \mu = 0$ and $p_+ \sim p_-$. Therefore, in this regime Eq.~\eqref{OhmAp} reduces to
\begin{eqnarray}\label{InertialOhmAp}
&& U_\nu F^{\mu\nu} = \eta \left[ {J^\mu - \rho '_e  (1 + \Theta) U^\mu} \right]  \nonumber\\
&& \qquad + \frac{1}{4en}\nabla_\nu\left[\frac{\xi h}{ne}\left(U^\mu J^\nu+J^\mu U^\nu-\frac{\Delta\mu}{ne}J^\mu J^\nu\right)\right]  \, , \nonumber\\
&&
\end{eqnarray}
where all non-ideal terms are displayed in the right-hand side.

The above equations naturally incorporate the effects of the spacetime curvature in the plasma dynamics. However, for our purposes, it is more effective to represent these equations by expressing them in the $3+1$ formalism \cite{TM_82,Thorne86,Zhang89}. Indeed, in this way, spacetime curvature effects can be displayed explicitly in a set of vectorial equations. Adopting the $3+1$ formalism, the line element becomes 
\begin{equation}
d{s^2} = g_{\mu \nu} d{x^\mu}d{x^\nu} =  - {\alpha ^2}d{t^2} + \sum\limits_{i = 1}^3 {{{({h_i}d{x^i} - \alpha {\beta^i}dt)}^2}} \, ,
\end{equation}
where 
\begin{equation}
\alpha= \left[ h_0^2 + \sum_{i = 1}^3 (h_i\omega _i)^2 \right]^{1/2} \, , \qquad \beta^i= \frac{{h_i}{\omega _i}}{\alpha}  \, ,
\end{equation}
are the lapse function and  the shift vector, respectively, while 
\begin{equation}
h_0^2=-g_{00} \, , \quad h_i^2=g_{ii} \, , \quad h_i^2 \omega_i = -g_{i0} = -g_{0i} \, ,
\end{equation}
are the non-zero components of the metric.

The plasma vectorial equations can be better understood introducing a locally nonrotating frame called ``zero-angular-momentum-observer'' (ZAMO) frame \cite{Bardeen_1972}. This frame offers the advantage of having a locally Minkowskian spacetime. Indeed, in this case, the line element can be simply written as 
\begin{equation}
d{s^2} =  - d{{\hat t}^2} + \sum\limits_{i = 1}^3 {{{(d{{\hat x}^i})}^2}}  = {\eta _{\mu \nu }}d{{\hat x}^\mu }d{{\hat x}^\nu } \, ,
\end{equation}
where
\begin{equation}
d\hat t = \alpha dt \, , \qquad d{{\hat x}^i} = {h_i}d{x^i} - \alpha {\beta^i}dt \, .
\end{equation}
Notice that here and in the following, quantities observed in the ZAMO frame are denoted with hats. A careful derivation of the generalized GRMHD equations in the ZAMO frame can be found in Ref.~\cite{Koide_2010}. Here, for the sake of compactness, we reproduce only the relevant ones for our study. The first one is the continuity equation \eqref{contAp}, which can be written as
\begin{equation}\label{contZAMO}
\frac{\partial(\gamma n)}{\partial t}+\frac{1}{h_1h_2h_3}\sum_j\frac{\partial}{\partial x^j}\left[\frac{\alpha h_1h_2h_3}{h_j}\gamma n\left(\hat v^j+\beta^j\right)\right]=0\, ,
\end{equation}
where $\hat v$ is the velocity observed in the ZAMO frame, and $\gamma  = {(1 - {\hat v^2})^{ - 1/2}}$ is its corresponding Lorentz factor (latin indices are used for space components).
Similarly, the spatial components of the generalized momentum equation \eqref{MomAp} can be written as
\begin{eqnarray}\label{momZAMO}
\frac{\partial \hat P^i}{\partial t}&=&-\frac{1}{h_1h_2h_3}\sum_j\frac{\partial}{\partial x^j}\left[\frac{\alpha h_1h_2h_3}{h_j}\left(\hat T^{ij}+\beta^j \hat P^i\right)\right]\nonumber\\
&&-(\epsilon+\gamma \rho)\frac{1}{h_i}\frac{\partial\alpha}{\partial x^i}-\sum_j \sigma_{ji}\hat P^j\nonumber\\
&&+\sum_j \alpha\left[G_{ij}\hat T^{ij}-G_{ji}\hat T^{jj}+\beta^j\left(G_{ij}\hat P^i-G_{ji}\hat P^j\right)\right]\, ,\nonumber\\
&&
\end{eqnarray}
with
\begin{equation}\label{PT1}
\hat P^i = h\gamma^2 \hat v^i+\frac{h\xi}{4 n^2e^2}\hat J^i\hat J^0+\sum_{j,k}\varepsilon_{ijk}\hat E_j \hat B_k \, ,
\end{equation}
\begin{equation}\label{}
\epsilon = h\gamma^2+\frac{h\xi}{4e^2n^2}(\hat J^0)^2-p-\rho\gamma+\frac{1}{2}\left(\hat B^2+\hat E^2\right) \, ,
\end{equation}
\begin{eqnarray}\label{PT2}
\hat T^{ij}&=&p\delta^{ij}+h\gamma^2\hat v^i\hat v^j+\frac{h\xi}{4e^2n^2}\hat J^i \hat J^j\nonumber\\
&&+\frac{1}{2}\left(\hat B^2+\hat E^2\right)\delta^{ij}-\hat B_i\hat B_j-\hat E_i\hat E_j \, .
\end{eqnarray} \\
Here, $\hat E_j$ and $\hat B_k$ are the electric and magnetic fields measured in the ZAMO frame, $\varepsilon_{ijk}$ is the Levi-Civita symbol, and
\begin{eqnarray}\label{ecGsigma}
G_{ij}=-\frac{1}{h_ih_j}\frac{\partial h_i}{\partial x^j}\, ,\qquad
\sigma_{ij}=\frac{1}{h_j}\frac{\partial(\alpha\beta^i)}{\partial x^j}\, .
\end{eqnarray}
 The generalized Ohm's law \eqref{InertialOhmAp} can also be written in the ZAMO frame. Its spatial components become
\begin{widetext}
\begin{eqnarray}\label{OhmZAMO}
&&\frac{\xi}{en}\frac{\partial}{\partial t}\left[\frac{h\gamma}{4en} \left(\hat J^i+\hat J^0\hat v^i\right)\right]=-\frac{1}{en h_1h_2h_3}\sum_j\frac{\partial}{\partial x^j}\left[\frac{\alpha h_1h_2h_3}{h_j}\left(\hat K^{ij}+\frac{h\xi\gamma}{4en}\beta^j\left(\hat J^i+\hat J^0\hat v^i\right)\right)\right]\nonumber\\
&&\qquad-\frac{h\gamma\xi\hat J^0}{2e^2n^2 h_i}\frac{\partial\alpha}{\partial x^i}
+\frac{\alpha}{en}\sum_j \left[G_{ij}\hat K^{ij}-G_{ji}\hat K^{jj}+\frac{h\xi\gamma}{4en}\beta^j\left(G_{ij}(\hat J^i+\hat J^0\hat v^i)-G_{ji}(\hat J^j+\hat J^0\hat v^j)\right)\right]\nonumber\\
&&\qquad -\frac{h\xi\gamma}{4e^2n^2}\sum_j \sigma_{ji}(\hat J^j+\hat J^0\hat v^j)+\alpha\gamma \hat F_{i0}+\alpha\gamma\hat v^j\hat F_{ij}-\alpha\eta \left[ \hat J^i - \rho '_e  (1 + \Theta) \gamma \hat v^i \right]\, ,
\end{eqnarray}
\end{widetext}
 where 
\begin{equation}
\hat K^{ij}=\frac{h\xi\gamma}{4en}\left[\hat v^i\hat J^j+\hat v^j\hat J^i\right]\, .
\end{equation}
Finally, we rewrite Maxwell's equations \eqref{MaxGeneralCurvedAp} in the ZAMO frame.  The two constraint equations become
\begin{equation}\label{Maxw1ZAMO}
\sum_j\frac{\partial}{\partial x^j}\left(\frac{h_1h_2h_3}{h_j}\hat B_j\right)=0\, ,
\end{equation}
\begin{equation}\label{Maxw3ZAMO}
\frac{1}{h_1h_2h_3}\sum_j\frac{\partial}{\partial x^j}\left(\frac{h_1h_2h_3}{h_j}\hat E_j\right)=\hat J^0\, ,
\end{equation}
while the Ampere-Maxwell equation can be written as
\begin{eqnarray}\label{Maxw2ZAMO}
&&\alpha\hat J^i+\alpha\hat J^0\beta^i+\frac{\partial \hat E_i}{\partial t}\nonumber\\
&&=\frac{h_i}{h_1h_2h_3}\sum_{j,k}\varepsilon^{ijk}\frac{\partial}{\partial x^j} \Bigg[ \alpha h_k \Bigg( \hat B_k+\sum_{l,m}\varepsilon_{klm}\beta^l \hat E_m \Bigg) \Bigg] \, ,\nonumber\\
&&
\end{eqnarray}
and Faraday's law as
\begin{eqnarray}\label{Maxw4ZAMO}
&&\frac{\partial \hat B_i}{\partial t}\nonumber\\
&&=\frac{-h_i}{h_1h_2h_3}\sum_{j,k}\varepsilon^{ijk}\frac{\partial}{\partial x^j} \Bigg[ \alpha h_k \Bigg( \hat E_k-\sum_{l,m}\varepsilon_{klm}\beta^l \hat B_m \Bigg) \Bigg] \, .\nonumber\\
&&
\end{eqnarray}

In the following, we use these equations written in the ZAMO frame to analyze the magnetic reconnection process around rotating black holes. Our purpose is to investigate if, and how, the spacetime curvature produced by the black hole affects the reconnection mechanism.

\section{Spacetime and Reconnection Layer Configuration}

In this section, we specify the spacetime around the rotating black hole, $(x^0, x^1, x^2, x^3) = (t, r, \theta, \phi)$, as well as the configuration of the magnetic reconnection layer. The metric of a rotating black hole with mass $M$ and angular momentum $J$ is given by the Kerr metric \cite{Weinberg_72}, for which 
\begin{equation}\label{}
h_0 = {\left( {1 - \frac{{2{r_g}r}}{\Sigma }} \right)^{1/2}}  \, ,\quad h_1 = {\left( {\frac{\Sigma }{\Delta }} \right)^{1/2}} \, ,
\end{equation}
\begin{equation}\label{}
h_2 = {\Sigma}^{1/2} \, ,\quad h_3 = {\left( {\frac{A}{\Sigma}} \right)^{1/2}} \sin \theta \, ,
\end{equation}
\begin{equation}\label{}
\omega_1 = \omega_2 = 0\, ,\quad \omega_3 = \frac{{2r_g^2ar}}{\Sigma} \, .
\end{equation} 
Here, $r_g = GM$ is the gravitational radius, with $G$ denoting the gravitational constant, and $a = J/J_{\max}\leq 1$ is the rotation parameter, with $J_{\max} = GM^2$ indicating the angular momentum of a maximally rotating black hole. Furthermore, $\Sigma $, $\Delta$ and $A$, which have been introduced for brevity, are defined as  
\begin{equation}\label{}
\Sigma  = {r^2} + {\left( {a{r_g}} \right)^2}{\cos ^2}\theta \, ,
\end{equation}
\begin{equation}\label{}
\Delta  = {r^2} - 2{r_g}r + {\left( {a{r_g}} \right)^2} \, ,
\end{equation}
\begin{equation}\label{}
A = \left[ {{r^2} + {{\left( {a{r_g}} \right)}^2}} \right]^2 - \Delta {\left( {a{r_g}} \right)^2}{\sin ^2}\theta \, .
\end{equation}
Finally, the lapse function and the shift vector are given by
\begin{equation}\label{Maxw3ZAMO}
\alpha = {\left( {\frac{{\Sigma \Delta }}{A}} \right)^{1/2}}  \, ,  \qquad  \beta^j=\beta^\phi \delta^{j\phi}  \, ,
\end{equation}
where $\beta^\phi=h_3\omega_3/\alpha$ is a measurement of the rotation of the Kerr spacetime, in which $\beta^jG_{ij} = 0$. 
The radius of the event horizon can be obtained by setting $\alpha=0$, which leads to $r_H = r_g (1 + \sqrt {1 - {a^2}} \,)$.

\begin{figure}[]
\begin{center}
\includegraphics[width=8.5cm]{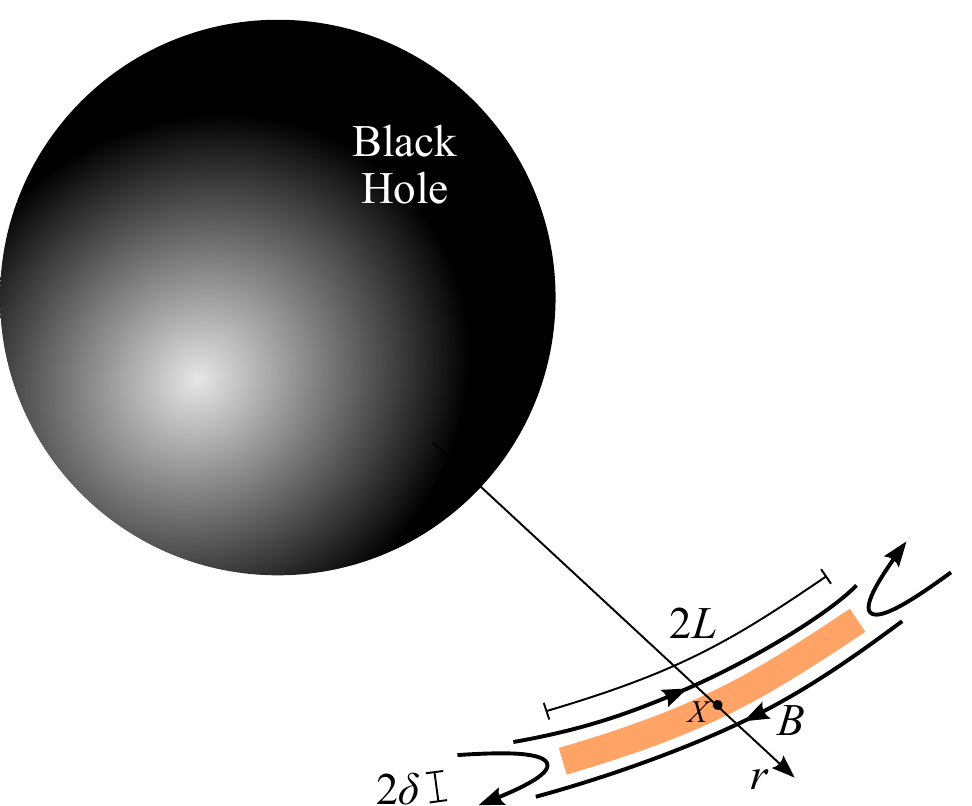}
\end{center}
\caption{Sketch of a reconnection layer in the azimuthal direction around a rotating black hole. The shaded orange area represents a narrow ($\delta\ll L$) magnetic diffusion region.}
\label{fig1}
\end{figure}

Magnetic reconnection layers can form in different locations around the rotating black hole. Here, we consider the typical situation where the reconnection layer is in the equatorial plane, $\theta = \pi/2$, or very close to it. In particular, we examine the case in which the magnetic diffusion region is oriented in the azimuthal direction, as shown in Fig.~\ref{fig1}. This orientation is indeed commonly found in numerical simulations (e.g., Refs. \cite{Penna2010,McKinney2012,Ball2017}).  Finally, we assume that the reconnection process occurs in a stable orbit around the rotating black hole \cite{abramo,Koide2011,Tursunov2016}, in such a way that the plasma is supported against the black hole gravity.

We adopt a Sweet-Parker-like approach \cite{lyu,luca1,asenjoGRLuca} in order to evaluate the reconnection rate and other quantities related to the magnetic reconnection layer. This approach is suitable to study magnetic reconnection within narrow quasi-two-dimensional current sheets, i.e. $\delta \ll L$, under quasi-stationary conditions, i.e. $\partial_t \approx 0$. The latter condition is satisfied not only in steady-state, but also at the time of maximum reconnection rate. The reconnection layer is implicitly assumed to be stable to the plasmoid instability \cite{Comisso_2016}. If this is not the case, the global reconnection layer would be replaced by a chain of plasmoids of different sizes separated by smaller current sheets \cite{Shibata2001}, but our analysis would still be valuable for understanding the properties of the basic  current sheets composing the global reconnection layer \cite{Huang_2010,ULS_2010,Comisso2015,Comisso2016,Belob2017}.

\section{Reconnection layer analysis}

In our reconnection layer analysis, we first discuss the case in which the current density in the reconnection layer is in the $\theta$-direction, and then we extend the calculation to investigate if a reconnection layer having current density in a different direction could be characterized by a different reconnection rate.

\subsection{Poloidally-oriented current density}

In the azimuthal configuration \cite{asenjoGRLuca}, the reconnecting magnetic field is in the $\phi$-direction. We indicate with ${\hat B}_{\rm{in}}$ its magnitude in the ZAMO frame. In the diffusion region, the radial velocity $\hat v^r$ vanishes at the neutral line, where $\hat B^\phi = 0$. Similarly, at the neutral line $\hat J^r = 0 \approx \hat J^\phi$, and $\hat J^0=0$ by quasi-neutrality, implying that the charge density observed by the local center-of-mass frame of the plasma vanishes. 
We assume $\hat v^\theta \approx 0$, $\hat B^\theta \approx 0$, and that spatial variations of the fields with respect to $\theta$
are negligible. The same configuration has been adopted by Koide and Arai \cite{Koide_2010K} to examine the possibility of energy extraction from a rotating black hole. Indeed, magnetic reconnection can redistribute the plasma angular momentum to yield negative energy at infinity of the plasma, making it an interesting alternative to the well-known Penrose \cite{Penrose69,PF71} and Blandford-Znajek \cite{BZ77} processes.

In order to obtain the outflow velocity of the accelerated plasma through the magnetic reconnection process, we use the momentum equation \eqref{momZAMO} written in the $3+1$ formalism. Evaluating it along the neutral line, we find 
\begin{equation}\label{momI1}
\frac{1}{h_1h_2h_3}\frac{\partial}{\partial\phi}\left[\alpha h_1h_2\left(\hat T^{\phi\phi}+\beta^\phi\hat P^\phi\right)\right] = \alpha G_{\phi j}\hat T^{\phi j}\, ,
\end{equation}
 where the relevant components of $\hat P$ and  $\hat T$, which can be found from Eqs. \eqref{PT1} and \eqref{PT2}, are
\begin{equation}\label{}
\hat P^\phi = h\gamma^2 \hat v^\phi \, ,
\end{equation}
\begin{equation}\label{}
\hat T^{\phi j} = p{\delta ^{\phi j}} + h{\gamma ^2}{\hat v^\phi }{\hat v^j} + \frac{1}{2} \big( {{\hat B}_{\rm{in}}^2 - \hat E_\theta ^2} \big){\delta ^{\phi j}} - {\hat B_\phi }{\hat B_j} \, .
\end{equation}
Using Eq. \eqref{ecGsigma} we find that $G_{\phi j}\hat T^{\phi j} = 0$. Thus, Eq.~\eqref{momI1} reduces to
\begin{equation}\label{momI2}
\frac{\partial}{\partial\phi} \Bigg[ h\gamma^2\hat v^\phi\left(\hat v^\phi+\beta^\phi\right)-\frac{{\hat B}_{\rm{in}}^2}{2}-p \Bigg] = 0 \, .
\end{equation}
The magnetohydrodynamic enthalpy density $h$ can be obtained from the equation of state of each fluid. In thermal equilibrium we have \cite{Chandra1938,Synge1957}
\begin{equation}\label{}
h = mn\frac{{{K_3}(m/{k_B}T)}}{{{K_2}(m/{k_B}T)}}  \, ,
\end{equation}
where $K_2$ and $K_3$ are the modified Bessel functions of the second kind of orders two and three, respectively. Hence, for a relativistically hot plasma ${h} \approx 4p$. At this point, we are left with the evaluation of the pressure at the center of the current sheet. This can be done from the assumption of pressure balance across the layer, which gives us $p \approx {{\hat B}_{\rm{in}}^2}/2$. Substituting this relation into Eq. \eqref{momI2}, we can finally evaluate the outflow velocity. Thus, from the integration of Eq. \eqref{momI2}, we conclude that
\begin{equation}\label{outflowMR}
\gamma_{\rm{out}} {\hat v}_{\rm{out}} \approx 1/\sqrt{2} \, ,
\end{equation}
implying that the outflow velocity and its Lorentz factor are both of order of unity, as was also shown for current sheet configurations around Kerr black holes in the absence of thermal-inertial effects \cite{asenjoGRLuca}. 
Note that this conclusion is not affected by the rotation of the black hole. Indeed, when evaluating this effect along the neutral line, at the $X$-point, the $\beta^\phi  v^\phi$ contribution vanishes, while at the outflow point $\beta^\phi$ is at best of the same order as $v^\phi$, which do not modify our order of magnitude approximation.

We proceed with the calculation of the reconnected magnetic field at the ouflow point by assuming magnetic flux conservation through the current sheet. Using Eq. \eqref{Maxw1ZAMO} we end up with
\begin{equation} \label{flux_cons}
\hat B_r |_{\rm{out}} \approx \left. {\left( {\frac{{r{h_1}}}{{{h_3}}}} \right)} \right|_{\rm{out}} \frac{\delta }{L} {\hat B}_{\rm{in}} \, ,
\end{equation} 
where the symbol $|_{\rm{out}}$ indicates that the relevant quantities are evaluated at the outflow point of the reconnection layer. Similarly, we can express the inverse-aspect-ratio of the reconnection layer by assuming steady-state flow flux conservation. Using Eq.~\eqref{contZAMO}, if the inflow flux $\gamma_{\rm in} \hat v_{\rm in}$ balances the outflow flux $\gamma_{\rm{out}} \hat v_{\rm{out}}$, we are led to the relation
\begin{equation} \label{deltaLest}
\frac{\delta}{L} \approx {\left. {\left( {\frac{{{h_3}}}{{{h_1}r}}} \right)} \right|_{\rm{out}}}\,\frac{{\gamma}_{\rm in}\hat v_{\rm in}}{{\gamma}_{\rm{out}} \hat v_{\rm{out}}}  \, ,
\end{equation}
where we have considered Eq.~\eqref{flux_cons} in the estimation. 

The evaluation of the generalized Ohm's law \eqref{OhmZAMO} constitutes the last step required to estimate the velocity $\hat v_{\rm in}$ that measures the rate at which the magnetic flux undergoes the reconnection process. We consider separately the inner region, where magnetic diffusion occurs, from the outer region, where the plasma moves with a transport velocity that preserves the magnetic connections between plasma elements \cite{Newcomb,pegoraroEPJ,asenjoComissoCon,AsenjoComissoPRD17,Beken78,asenjo2015}. Since $\partial_t \approx 0$ and $\partial_\theta \approx 0$, from Eq. \eqref{Maxw4ZAMO} we have ${\partial_r} (\alpha {h_2}{{\hat E}_\theta }) = 0$ along the inflow line passing through the $X$-point. Because of the smallness of the current layer width $\delta$, this implies that ${\hat E}_\theta |_{\rm{in}} \approx {\hat E}_\theta |_X$. Therefore, we can match the electric field ${\hat E}_\theta$ at the $X$-point and the inflow point in order to obtain the inflow velocity $\hat v_{\rm in}$. In the current sheet, the generalized Ohm's law \eqref{OhmZAMO} in the $\theta$-direction is
\begin{eqnarray}\label{OhmGeneral_theta}
&&\frac{1}{{en{h_1}{h_2}{h_3}}}\sum\limits_j {\frac{\partial }{{\partial {x^j}}}} \left[ {\frac{{\alpha {h_1}{h_2}{h_3}}}{{{h_j}}}\left( {{{\hat K}^{\theta j}} + \frac{{h\xi \gamma }}{{4en}}{\beta ^j}{{\hat J}^\theta }} \right)} \right] \nonumber\\
&& - \frac{\alpha }{{en}}\sum\limits_j {\left[ {{G_{\theta j}}{{\hat K}^{\theta j}} - {G_{j\theta }}{{\hat K}^{jj}} + \frac{{h\xi \gamma }}{{4en}}{\beta ^j}\left( {{G_{\theta j}}{{\hat J}^\theta } - {G_{j\theta }}{{\hat J}^j}} \right)} \right]} \nonumber\\
&& + \frac{{h\xi \gamma }}{{4{e^2}{n^2}}}\sum\limits_j {{\sigma _{j\theta }}} {{\hat J}^j} - \alpha \gamma {{\hat F}_{\theta 0}} - \alpha \gamma {{\hat v}^j}{{\hat F}_{\theta j}} + \alpha \eta {{\hat J}^\theta } = 0 \, ,
\end{eqnarray}
At the inflow point, where all non-ideal terms can be neglected, the Ohm's law simply gives
\begin{equation}\label{Ohminflow}	
{\hat E}_\theta |_{\rm{in}} \approx \hat v_{\rm in} {\hat B}_{\rm{in}} \, .
\end{equation}
On the other hand, at the $X$-point, where the plasma velocity vanishes, the evaluation of Eq.~\eqref{OhmGeneral_theta} leads us to
\begin{equation}\label{OhmXpoint}	
{\hat E}_\theta |_X \approx (\eta+\Lambda) \hat J^\theta |_X  \, ,
\end{equation}
where 
\begin{equation} \label{Eff_Res}
\Lambda=\frac{\xi h}{4n^2e^2 L}\left.\frac{r}{h_3} \gamma\hat v^\phi\right|_{\rm{out}}  \approx \frac{\xi h}{4n^2e^2 L}\left.\frac{r}{h_3} \right|_X \, .
\end{equation}
It is possible to regard $\Lambda$ as an ``effective resistivity'' given by thermal-inertial-curvature effects. Note that $r_X <\left.{h_3} \right|_X$ in general. Therefore, this new effective resistivity is smaller than the one obtained in the flat spacetime limit \cite{luca1}. For small black hole rotation we can adopt the approximation
\begin{equation}
\left.\frac{r}{h_3} \right|_X \approx 1-\frac{a^2 r_g^2}{2 r_X^2} \, ,
\end{equation}
which clearly shows that this effective resistivity decreases as the reconnection layer  becomes closer to the black hole.

In Eq. \eqref{OhmXpoint}, the current density at the $X$-point can be estimated from Eq. \eqref{Maxw2ZAMO} as
\begin{equation}\label{}
{{\hat J}^\theta} |_X \approx  - \left.\frac{1}{h_1} \right|_{X} \frac{{\hat B}_{\rm{in}}}{\delta} \, .
\end{equation} 
Therefore, the matching of the expressions for ${\hat E}_\theta |_{\rm{in}}$ and ${\hat E}_\theta |_X$ leads us to the conclusion that 
\begin{equation} \label{ReconnectionrateGRazim}
\hat v_{\rm in}  \approx  \sqrt{ {\left.\frac{r}{h_3} \right|_X} \left( {\frac{1}{S} + \frac{\Lambda }{L}} \right) } \, ,
\end{equation}
where $S = L/\eta \gg 1$ indicates the Lundquist number, which represents the dimensionless ratio between the Alfv\'en wave crossing timescale and the resistive diffusion timescale. This expression shows that the collisionless effects retained in $\Lambda$ increase the reconnection rate with respect to the purely collisional case \cite{asenjoGRLuca}. However, the net effect is smaller compared to the thermal-inertial effects in flat spacetime \cite{luca1}, because the spacetime curvature induced by the rotating black hole acts as to increase the aspect ratio of the reconnection layer.

It must be noted that $S$ is a remarkably large number for the highly conducting plasmas of interest here (it can be as large as $10^{20}$ or even greater \cite{Loeb2007}). Consequently, collisionless effects are crucial to substain very high reconnection rates in the vicinity of black holes. Note that the reconnection rate given by Eq.~\eqref{ReconnectionrateGRazim} can be high even for $S \to \infty$ because of the thermal-inertial-curvature effects contained in the expression \eqref{Eff_Res}. More generally, the thermal-inertial-curvature effects dominate over the collisional ones when $\Lambda/L \gg 1/S$. This is indeed expected to be the case for the hot tenuous plasmas surrounding black holes. 

While until now we have not made any assumption about the constituent particles of the plasma, which are specified though $\xi$, it is useful to consider this new collisionless regime in a more definite way. For an electron-ion plasma, thermal-inertial-curvature effects dominate when 
\begin{equation}
\frac{h m_-}{n^2e^2 m_+ L^2}\left.\frac{r}{h_3} \right|_X \approx \frac{f}{\omega_{pe}^2 L^2}\left.\frac{r}{h_3} \right|_X \gg \frac{1}{S} \, ,
\end{equation}
where we have indicated with $\omega_{pe}$ the plasma frequency and with $f=K_3(m/k_B T)/K_2(m/k_B T)$ the relativistic thermal factor. In this regime, the reconnection rate becomes
\begin{equation}\label{ReconnectionrateGRazimEI}
\hat v_{\rm in}^{e-i} \approx {\left.\frac{r}{h_3} \right|_X}  \sqrt{f} \frac{\lambda_e}{L} \, ,
\end{equation}
while the reconnection layer width turns out to be
\begin{equation}\label{LayerWidthGRazimEI}
{{\delta}^{e-i}} \approx {\left.\frac{1}{h_1} \right|_X}  \sqrt{f} \lambda_e \, ,
\end{equation}
with $\lambda_{e}$ indicating the electron skin depth. On the other hand, for a pair plasma, the thermal-inertial-curvature effects dominate when 
\begin{equation}
\frac{h}{4 n^2e^2 L^2}\left.\frac{r}{h_3} \right|_X = \frac{f}{2 \omega_{pe}^2 L^2}\left.\frac{r}{h_3} \right|_X \gg \frac{1}{S} \, .
\end{equation}
In this case, the reconnection rate and the reconnection layer width become 
\begin{equation}\label{ReconnectionrateGRazimPAIR}
\hat v_{\rm in}^{pair} \approx {\left.\frac{r}{h_3} \right|_X}  \sqrt{\frac{f}{2}} \frac{\lambda_e}{L}  \, , \quad  {\delta}^{pair} \approx {\left.\frac{1}{h_1} \right|_X}  \sqrt{\frac{f}{2}} \lambda_e  \, .
\end{equation}
When considering the flat spacetime limit, $\left. h_3 \right|_X=r_X$, these formulas reduce to the ones obtained in Ref. \cite{luca1} (the factor $\sqrt{2}$ difference appears because here $f$ is defined such that $h = h_+ + h_- \approx 2 m_- n f$ for the pair plasma case), with reconnection rates that can be even larger than $\hat v_{\rm in}^{pair} \sim 0.1$  \cite{Zenitani2009,Bessho2012,Liu2015,ComissoJPP16}.

\subsection{Radially-oriented current density}

To explore if the direction of the current density in the reconnection layer could lead to a different reconnection rate, here we consider a similar configuration, but with current density in the $r$-direction. At the neutral line we have $\hat J^\theta = 0 \approx \hat J^\phi$, as well as $\hat J^0=0$. Again, we assume $\hat v^r \approx 0$, $\hat B^r \approx 0$, and that spatial variations of the fields with respect to the radial distance are negligible, i.e. $\partial_r \approx 0$. 

In the evaluation of the outflow velocity, the only difference with the above analysis is that now the electric field $\hat E_\theta$ in the expression for $\hat T^{\phi j}$ is replaced by the $\hat E_r$ component. This, however, does not play a role in the momentum equation along the neutral line, Eq. (\ref{momI1}), because at the neutral line ${\partial_\phi} \hat E_r=0$. Therefore, employing the same approximations adopted before, we obtain again $\gamma_{\rm{out}} {\hat v}_{\rm{out}} \approx 1/\sqrt{2}$. On the other hand, for this configuration, magnetic flux conservation through the current sheet yields the reconnected magnetic field
\begin{equation}\label{outflowMagneticF_2}
\hat B_\theta |_{\rm{out}} \approx {\left.\frac{r}{h_3}\right|_{\rm{out}}}\frac{\delta}{L} {\hat B}_{\rm{in}}\, ,
\end{equation}
where we have estimated $\delta\phi \approx L/r_{\rm{out}}$ and $\delta \theta \approx \delta/r_{\rm{out}}$. The expression for the inverse-aspect-ratio of the reconnecting current sheet is also slightly different. Indeed, from the continuity equation \eqref{contZAMO} we can obtain
\begin{equation}\label{deltaLest_2}
\frac{\delta}{L} \approx {\left. \frac{h_3}{r} \right|_{\rm{out}}} \,\frac{{\gamma}_{\rm{in}}\hat v_{\rm{in}}}{{\gamma}_{\rm{out}}\hat v_{\rm{out}}} \, .
\end{equation}

Following the same procedure adopted before, we use the generalized Ohm's law \eqref{OhmZAMO} to complete the relations needed to calculate the rate at which the magnetic flux is transported across the diffusion region. Indeed, since $\partial_t \approx 0$ and $\partial_r \approx 0$, the electric field ${{\hat E}_r}$ is uniform along the line passing the inflow and $X$ points, allowing us to employ the standard matching procedure by means of the generalized Ohm's law. In the current sheet, Eq. \eqref{OhmZAMO} in the radial direction reduces to
\begin{eqnarray}\label{OhmGeneral_r}
&&\frac{\xi}{\alpha e n h_1h_2h_3}\frac{\partial}{\partial\phi}\left[\frac{\alpha h_1h_2 h\gamma}{4 e n}\hat J^r(\hat v^\phi+\beta^\phi)\right]\nonumber\\
&&\frac{h\xi\gamma}{4e^2n^2 h_1 h_j}\left(\frac{\partial h_1}{\partial x^j}\hat v^j\hat J^r-2\frac{\partial h_j}{\partial r}\hat v^j\hat J^j\right) + \eta\hat J^r\nonumber\\
&& -\gamma \hat F_{r0} -\gamma\hat v^\theta \hat F_{r\theta} -\gamma\hat v^\phi\hat F_{r\phi} = 0 \, .
\end{eqnarray}
At the $X$-point, this equation yields
\begin{equation}\label{Ohmoutflow2}
{\hat E}_r |_{X} \approx \left(\eta+ \Lambda \right) {\hat J}^r |_{X} \, ,
\end{equation}
which is very similar to the reconnection electric field obtained before for a $\theta$-oriented current density. The only difference is that now, from Eq. \eqref{Maxw2ZAMO}, the current density at the $X$-point is simply $\hat J^r |_{X}  \approx {{\hat B}_{\rm{in}}}/{\delta}$.
Matching Eq.~\eqref{Ohmoutflow2} with the electric field expression at the inflow point, ${\hat E}_r |_{\rm{in}} \approx \hat v_{\rm in} {\hat B}_{\rm{in}}$, with the help of Eqs.~\eqref{outflowMagneticF_2} and \eqref{deltaLest_2}, we obtain the inflow plasma velocity in $\theta$-direction, whose final expression is $\hat v_{\rm in} \approx  \sqrt{{\left. (r/{h_3}) \right|_X} (1/S+\Lambda/L)} $, as for the $\theta$-oriented current density. Therefore, we conclude that the orientation of the current density does not have a significant impact on the reconnection rate.

\section{Conclusions}

In this paper, we have studied the magnetic reconnection process in curved spacetime due to the presence of a rotating black hole. By performing a Sweet-Parker-like analysis for azimuthal reconnection layers, we have calculated the reconnection rate as well as other important quantities such as the width of the reconnection layer, the reconnected magnetic field, and the outflow velocity of the plasma accelerated through the magnetic reconnection process. This analysis has allowed us to obtain the first order effects induced by the gravitational field of a rotating black hole. In particular, we have shown that the spacetime curvature due to the black hole rotation acts to decrease the reconnection rate in azimuthal reconnection layers.

The analysis presented here extends our recent Letter \cite{asenjoGRLuca} to include also plasma collisionless effects, which couple to gravity and have the net effect of enhancing the rate at which the magnetic flux is transported toward the reconnection $X$-point. Indeed, we have shown that thermal-inertial-curvature effects cause a broadening of the reconnection layer, which, in turn, enables higher reconnection rates. This has been shown for both electron-ion and pair plasmas, and can be understood in terms of an ``effective resistivity'' $\Lambda$ that limits the response of the electrons (or electrons and positrons) to the reconnection electric field. The thermal-inertial-curvature resistivity $\Lambda$ depends on the thermal factor $f$, the plasma frequency $\omega_{pe}$, and the curvature-related ratio $ {\left. (r/{h_3}) \right|_X}$. Therefore, these effects can be very important in the hot tenuous plasmas surrounding black holes, where the condition $\Lambda/L \gg 1/S$ is expected to occur.

We observe that the ideas presented here about the interaction between gravity and magnetic reconnection may have a much broader impact. If elaborated further, they may indicate whether magnetic reconnection could be an efficient mechanism of energy extraction from rotating black holes. Besides, they could be adopted to understand how gravitational effects can influence magnetic energy release rates close to neutron stars. 

Finally, one might wonder if pure-relativistic effects can play the role of an effective resistivity in more general configurations than the ones proposed here. These kind of effects were qualitatively explored by Koide in Ref.~\cite{Koide_2010}, and they correspond, for example, to the terms $-({h\gamma\xi\hat J^0}/{2e^2n^2 h_i})({\partial\alpha}/{\partial x^i})$ or $({\alpha}/{en}) \sum\nolimits_j (G_{ij}\hat K^{ij}-G_{ji}\hat K^{jj})$ in the generalized Ohm's law \eqref{OhmZAMO}. In our model, all those terms vanish. This occurs because effective resistivities generated by these pure-relativistic terms are only possible in more complex configurations, as three-dimensional models. We will explore these ideas in forthcoming works.

\begin{acknowledgments}
It is a pleasure to acknowledge fruitful discussions with Gabriele Brambilla, Luis Lehner, Manasvi Lingam, Russell Kulsrud, and Alexander Tchekhovskoy. F.A.A. thanks Fondecyt-Chile  Grant No. 11140025.
L.C. is grateful for the hospitality of the Universidad Adolfo Ib\'a\~nez, where part of this work was done. 
\end{acknowledgments}

\end{document}